\documentclass[manuscript]{aastex}

\usepackage{amsmath}
\usepackage{subfig}
\def\fermi{{\it Fermi\/}}
\def\swift{{\it Swift\/}}

\def\xmm{{\it XMM-Newton\/}}

\def\lsim{\mathrel{\lower .85ex\hbox{\rlap{$\sim$}\raise
.95ex\hbox{$<$} }}}
\def\gsim{\mathrel{\lower .80ex\hbox{\rlap{$\sim$}\raise
.90ex\hbox{$>$} }}}
\newbox\grsign \setbox\grsign=\hbox{$>$}
\newdimen\grdimen \grdimen=\ht\grsign
\newbox\laxbox \newbox\gaxbox
\setbox\gaxbox=\hbox{\raise.5ex\hbox{$>$}\llap
     {\lower.5ex\hbox{$\sim$}}}\ht1=\grdimen\dp1=0pt
\setbox\laxbox=\hbox{\raise.5ex\hbox{$<$}\llap
     {\lower.5ex\hbox{$\sim$}}}\ht2=\grdimen\dp2=0pt

\shorttitle{A Point-Source Catalog of Unid \fermi\ Sources}
\shortauthors{Mirabal}

\begin{document}

\title{A Point-Source Catalog from \swift\ Observations of Unidentified 
\fermi\ Sources}

\author{N. Mirabal\altaffilmark{1,2}}

\altaffiltext{1}{Ram\'on y Cajal Fellow; Dpto. de F\'isica At\'omica, 
Molecular y Nuclear, Universidad Complutense de 
Madrid, Spain}
\altaffiltext{2}{Email: mirabal@gae.ucm.es}

\begin{abstract}

We present X-ray point source catalogs 
obtained from archival
\swift\ observations that have partially or fully covered
the  95\% confidence
error contour of unidentified \fermi\ sources. In total, 
21 out of 37 unidentified \fermi\ sources have been observed
by the X-Ray Telescope (XRT) on board the \swift\ observatory. Basic
properties such as position, positional errors, count
rates, and hardness ratios are derived for a total of 
18 X-ray point sources found in the
sample. From these detections, we discuss 
potential counterparts to 0FGL J0910.2$-$5044,
0FGL J1045.6$-$5937, 
0FGL J1115.8$-$6108, 0FGL 
J1231.5$-$1410, 0FGL J1326.6$-$5302, 0FGL J1604.0$-$4904,
and 0FGL J1805.3$-$2138. The catalog will assist observers  
planning future programs to identify 0FGL sources without obvious
counterparts. 
\end{abstract}


\section{Introduction}
Until recently, nearly the totality of $\gamma$-ray sources 
in the inner Galaxy  $l = 270^\circ - 90^\circ$
and $-30^\circ < b <  30^\circ$ remained without firm counterparts.
It has been long suspected  that low-latitude sources 
comprise a Galactic population that is either similar to the
already identified $\gamma$-ray pulsars \citep{ion}, or represent
an entirely new class of $\gamma$-ray emitters associated with the disk/bulge
population \citep{thompson,john}.   With improved
localizations and superb timing capabilities, \fermi\ has started to unveil
the likely culprits \citep{abdo1}. Among the
 sources reported 
in the Bright Gamma-ray Source List (0FGL), 26 include 
pulsars, supernova remnants, and pulsar wind nebulae \citep{roberts}
within the inner Galaxy. 
While we are certainly on our
way to solving this long-standing puzzle, the nature of approximately half 
of the 0FGL sources in the inner Galaxy remains elusive. 

It is likely that additional \fermi\ sources will be identified through
$\gamma$-ray timing. Others will require intensive
work at other wavelengths to produce plausible counterparts. Important 
challenges await for sources that fail to reveal themselves easily. 
Radio surveys will be of limited help 
for the potential analogs to the radio-quiet pulsars such as
 0FGL J0007.4+7303 \citep{halpern2,abdo3} 
and 0FGL J1836.2+5925 \citep{mirabal,reim,halpern1,abdo2}. 
In addition, extreme 
dust extinction and reddening close to the Galactic plane 
will be difficult to 
overcome in the optical. As a consequence, hard X-rays 
may provide the only opportunity to penetrate the Galactic barrier 
and gain access to $\gamma$-ray objects in the inner Galaxy. 

X-ray observations were crucial in recent efforts to
secure likely counterparts for a pair of unidentified \fermi\ sources
\citep{bassani,mirabal2}. 
In order to extend 
the X-ray efforts and to help foster multiwavelength collaborations,
we present X-ray point source catalogs derived from archival 
\swift\ observations of unidentified \fermi\ sources. The paper
is organized as follows. Section 2 describes the observations and data
reductions. Sections 3 and 4 provide details about individual
\fermi\ sources. Lastly, discussion and conclusions are presented in
Section 5.

\section{Observations and Data Reduction}
Data for unidentified \fermi\ sources 
were retrieved from the public \swift\ archive and comprise observations
obtained during 3 November 2005$-$5 August 2009. 
In this period, a 
total of 21 unidentified \fermi\ sources were partially 
or fully covered with the X-Ray Telescope (XRT) on board \swift\
\citep{gehrels}. Figure 1 shows the distribution of 
unidentified \fermi\ sources gathered from \citet{abdo1} and highlights
the \fermi\ contours covered by \swift\ observations. All 
observations were obtained with the XRT operated in 
photon counting (PC) mode. The details of \swift\ XRT observations of 
unidentified \fermi\ sources are summarized in Table 1. 

Source extraction to identify all significant X-ray sources within
the \fermi\ error contours was performed with 
{\it wavdetect}. Source positions and positional errors 
were derived using {\it xrtcentroid}.  X-ray counts (0.3--10 keV) 
were extracted from a circular
region with a 20 pixel radius (47\arcsec). The background was
extracted from an annulus with a 20 pixel (inner radius)
to 30 pixel (outer radius) around the source. Throughout, we 
used {\it XSELECT} to filter counts with grades 0--12. 
In fields without significant detections, we placed upper limits
to the presence of point source emission for the particular \swift\ 
observation. X-ray point sources detected within the  95\% confidence 
error contour of unidentified \fermi\ sources are listed in 
Tables 2$-$16. Figures that show the 95\% \fermi\ error contours
superimposed on smoothed \swift\ XRT images can be found
at our website\footnote{See http://www.gae.ucm.es/$\sim$mirabal/Unidentified.html}.
 
In addition to position and count rate, we also
computed the hardness ratio (HR) after separating
the 2--10 keV {\it hard} band (H)  count rate 
and the 0.3--2 keV {\it soft} band (S) count rate for each source. The
ratio itself was derived by adopting HR = $\frac{H - S}{H + S}$. As a
result, negative values of HR indicate softer X-ray sources. 
We caution that certain 
unidentified \fermi\ sources suffer from large Galactic \ion{H}{1}
column densities $N_{\rm H}$ (fifth column
of Tables 2--16) as derived from the 
nH tool\footnote{http://heasarc.gsfc.nasa.gov/cgi-bin/Tools/w3nh/w3nh.pl}. 
In such cases, 
the observed HR most likely underestimates the actual amount of photons in
the {\it soft} 0.3--2 keV band.

\section{Notes on Individual Objects}
\label{sec:notes}

{\bf 0FGL J0614.3$-$3330:} This source is listed as 3EG J0616$-$3310 
in the 3EG catalog \citep{hartman}. Within the 
the \fermi\ 95\% confidence error contour lies the relatively soft 
X-ray source 
Swift J0614.5$-$3332. An \xmm\ object consistent with this 
position was suggested by \cite{lap} 
as the counterpart of 3EG J0616$-$3310. There are no archival 
radio sources consistent with the X-ray position.

{\bf 0FGL J0910.2$-$5044:} \citet{abdo1} declare this source variable in
$\gamma$-rays. The X-ray source Swift J0910.9$-$5048 is likely associated
with a blazar \citep{sadler} and Galactic plane transient 
at $b = -1.\!^{\circ}8$ \citep{che08,lan08}.

{\bf 0FGL J1231.5$-$1410:} Detected as EGR J1231$-$1412 in the EGR catalog
\citep{casan}. 
The brightest source in the field, Swift J1231.1$-$1411,
lacks a radio counterpart in the NRAO VLA Sky Survey (NVSS) source catalog 
\citep{condon}. The source is soft in X-rays with a hardness ratio HR = 
$-$1.00,
{\it i.e.} all its X-ray photons are detected at energies $E < 2.0$ keV. 
Inspection of a Digitized Sky Survey image reveals a blank field at the
X-ray position down to a
conservative limit of $R > 20.1$. We argue that Swift J1231.1$-$1411 
meets the requirements of a potential neutron-star counterpart
to 0FGL J1231.5$-$1410.

{\bf 0FGL J1311.9$-$3419:} Listed as 3EG 
J1314$-$3431 in the third EGRET catalog.
\citet{sowards} proposed the radio source J1316$-$3338 as the
likely counterpart for the original EGRET error contour. However,
this association is now excluded by the Fermi localization. The sole
X-ray source in this region, Swift J1311.5$-$3418,  could be associated with
NVSS J131130.97$-$341810.5 detected with a flux density 13.8 mJy at 1.4 GHz.

{\bf 0FGL J1326.6$-$5302:} We find a single X-ray source, Swift J1326.8$-$5256,
within
the \fermi\ error contour. A corresponding radio source PMN J1326$-$5256
was proposed by \citet{mirabal2}
as the likely counterpart of  0FGL J1326.6$-$5302. PMN J1326$-$5256 
was also tentatively associated with unidentified EGRET source
3EG J1316$-$5244 
\citep{bignall}.

{\bf 0FGL J1413.1$-$6203:} Originally discovered by
COS B (catalog name 2CG 311$-$01) and later confirmed by EGRET (3EG) and
AGILE 1AGL J1412$-$6149 \citep{pitto}.  No X-ray sources brighter than
2.5 $\times 10^{-3}$ s$^{-1}$ are detected within the
\fermi\ error contour. PSR J1412$-$6145 and J1413$-$6141
were proposed as possible counterparts of the original
EGRET source \citep{torres}. However,
both of these pulsars appear to be excluded as counterparts
with the new LAT position.  The LAT contour still
embeds faint shell  
arcs possibly associated with the supernova remnant
G312.4$-$0.4  \citep{case, doherty}. However,
the main shell structure of the remnant lies outside the Fermi error contour.

{\bf 0FGL J1536.7$-$4947:} Swift J1536.2$-$4944 
is the only prominent X-ray source
in the \fermi\ 95\% confidence error contour. The
source is positionally consistent with an extended 
radio source MRC 1532$-$495A clearly visible
at 843 MHz \citep{jones}. A corresponding radio source PMNM 153234.4$-$493426
was also detected 
in the Green Bank 4.85 GHz northern sky survey carried out during
1986 November and 1987 October \citep{gregory}.

{\bf 0FGL J1604.0$-$4904:} Inside the \fermi\ error contour lies  
a single source Swift J1603.8$-$4904. The X-ray source is also detected 
in radio as PMN J1603$-$4904 at 4.85 GHz with a flux density of 954 mJy.
It appears to be an excellent blazar candidate. 

{\bf 0FGL J1634.9$-$4737:} No X-ray sources were detected in the 
\swift\ pointing down to 2.7                                                
$\times 10^{-3}$ s$^{-1}$. We note that the Soft Gamma Ray Repeater  
SGR 1627$-$41 is localized outside the \fermi\ error contour. 

{\bf 0FGL J1653.4$-$0200:} Previously detected as 3EG J1652$-$0223
by EGRET \citep{hartman}. A radio pulsar search of the
original EGRET error circle with the Parkes
64$-$m radio telescope failed to detect
a plausible counterpart \citep{crawford}.
Two X-ray sources Swift J1653.3$-$0158 and Swift J1653.9$-$0203
 are detected within the \fermi\
error circle.    

{\bf 0FGL J1805.3$-$2138:} Lies in the vicinity of the TeV source
HESS J1804$-$216 \citep{karga}. This field was
partially covered by Swift. One intriguing 
X-ray source Swift J1804.5$-$2140 is detected within
the \fermi\ error contour. \citet{karga} suggested that
the latter could be an accreting binary or a pulsar wind 
nebula possibly associated with HESS J1804$-$216. 
However, no convincing evidence for pulsations has been found. 

{\bf 0FGL J1813.5$-$1248:} Initially listed as unidentified in the original
0FGL release. The $\gamma$-ray pulsar 
PSR 1813$-$1246 was later revealed at this position \citep{abdo1}.
Swift J1813.4$-$1246 most likely corresponds to 
to the X-ray counterpart to this source. 
One puzzle regarding this source concerns the fact 
that 0FGL J1813.5$-$1248 is the only \fermi\ pulsar 
listed as variable in the 0FGL.  
If real, the $\gamma$-ray 
variability could be indicative of a multiple sources overlapping in the
same line of sight (i.e. a $\gamma$-ray pulsar 
and $\gamma$-ray blazar in the same region of the sky). Alternatively 
0FGL J1813.5$-$1248 may well be a very peculiar $\gamma$-ray pulsar.
  
{\bf 0FGL J1830.3+0617:} We find a single X-ray source, Swift J1830.1+0619, 
within 
the \fermi\ error contour. \citet{mirabal2} have identified
this blazar as the likely counterpart of  0FGL J1830.3+0617.

{\bf 0FGL J2001.0+4352:} Two sources lie within the
the \fermi\ error contour. Of these, Swift J2001.2+4352 
was proposed by \citet{bassani} as the counterpart
of 0FGL J2001.0+4352.

{\bf 0FGL J2027.5+3334:} No prominent X-ray sources brighter than
1.5  $\times 10^{-3}$ s$^{-1}$ are detected within
the \fermi\ 95\% confidence error contour.

{\bf 0FGL J2055.5+2540:} Two X-ray sources lie 
within the \fermi\ error circle. 
Neither Swift J2055.8+2546 nor Swift J2055.8+2540
have apparent counterparts at 1.4 GHz. 

{\bf 0FGL J2241.7$-$5239:} Not detected by EGRET or AGILE.
One significant source Swift J2241.5$-$5246 lies at the edge of
the \fermi\ 95\% confidence error contour. This X-ray source appears
to have a radio equivalent in the stamp image extracted
from the 843 MHz Sydney University Molonglo
Sky Survey \citep{bock}.

{\bf 0FGL J2302.9+4443:} One low significant X-ray source was
detected at edge of \fermi\ error
contour on 2009 February, 15. Intriguingly, Swift J2302.1+4445 was not
detected in a second \swift\ pointing conducted $\approx 16$ days later. 
The reality of this source requires further confirmation.

\section{Special Cases}

A keen reader will notice the omission
of three additional unidentified \fermi\ sources 
that have been covered by \swift\ namely 0FGL J1045.6$-$5937,
0FGL J1115.8$-$6108, and 0FGL J1746.0$-$2900. 
We have deliberately labeled these sources as special cases given the
complexity of said regions.

{\bf 0FGL J1045.6$-$5937:} First observed as $\gamma$-ray source
1AGL J1043$-$5931 by the AGILE satellite \citep{tavani}. The 
\fermi\ 95\% confidence error contour
contains the colliding wind binary Eta Carinae and the Carina nebula
(NGC 3372). This region has been observed by \swift\ and other
major X-ray missions in multiple
occasions. The $\gamma$-ray emission could
be produced due to the interaction of colliding winds associated with
the Eta Carinae binary system \citep{benaglia,reimer}. However, an 
alternative $\gamma$-ray emitter cannot be excluded. One interesting
candidate is the neutron star candidate XMM J104608.72$-$594306.5 
discovered by \citet{hama}.

{\bf 0FGL J1115.8$-$6108:} This field contains the starburst region NGC 3603
located in the Carina spiral arm. It has been observed by 
\swift\ in three
separate visits. The massive stellar population in this region
includes numerous OB and Wolf-Rayet stars. The reader is referred to
\citet{moffat} for a detailed analysis. Among the massive star population, 
the binary system WR43a stands out. \citet{moffat} argued that
the scatter observed in
some of the most luminous X-ray sources in NGC 3603 may indicate the
presence of additional colliding wind binaries. As a result, it is important
to investigate whether massive colliding wind binaries are driving the
$\gamma$-ray production associated with this source. Nevertheless,
alternative emitters cannot be ruled out with the current observations.

{\bf 0FGL J1746.0$-$2900:}  Possibly associated with the Galactic
Center region. This general area has been observed multiple times 
by \swift. Unfortunately, cataloguing and modeling the X-ray emission
is challenging. We refer the reader to \citet{muno} for a point-source 
catalog of this region.

\section{Discussion and Conclusions}
We have presented point-source catalogs and  data 
analyses of 24 \swift\ XRT individual
observations that have partially or fully covered
the error contours of unidentified \fermi\ sources. In total,
we have detected 18 X-ray point sources distributed 
over 15 unidentified \fermi\ error contours. For 
0FGL J1413.1$-$6203, 0FGL J1634.9$-$4737, and 0FGL J2027.5+3334
we are only able to derive upper limit for source detections. 
In addition
to positions and count rates, we have computed the hardness 
ratios of detected sources. 

With a uniform dataset at hand, 
it is critical to examine these X-ray point sources as 
 potential counterparts to
unidentified \fermi\ sources. We have advanced initial interpretations 
for a handful. However, 
additional efforts are required to further characterize
the nature of the remaining counterpart candidates. 
It is our hope that this catalog release will  
motivate multiwavelength collaborations and help plan future observational
programs.  
In particular, as shown here 
continued observations with \swift\ and other major 
X-ray observatories will be key in advancing the identification
of \fermi\ sources within the inner Galaxy. In order to ensure
quick access to newer versions of this catalog, we will
provide regular updates through our website 
\footnote{See http://www.gae.ucm.es/$\sim$mirabal/Unidentified.html}.

\acknowledgments
I thank the Spanish Ministry of Science
and Technology for support 
through a Ram\'on y Cajal fellowship. I acknowledge
illuminating correspondence with Jules Halpern and Michael T. Wolff. I 
also gratefully acknowledge Teddy Cheung, David Thompson, and
the rest of the \fermi\ team for promoting the early multiwavelength 
efforts.

\begin{deluxetable}{lcccc}
\tablewidth{0pt}
\tabletypesize{\tiny}
\tablecaption{Unidentified \fermi\ sources observed by \swift\ \label{fermitable}}
\tablehead{
\colhead{Name OFGL}       & \colhead{ObsID}                &
\colhead{Date}            & \colhead{Start Time}                &
\colhead{Exposure}          \\
%
%
\colhead{}                        & \colhead{}         &
\colhead{}     & \colhead{UT} & \colhead{\scriptsize (sec)}
}
\startdata
J0614.3$-$3330 & 00031375001  & 2009-03-16    & 06:45:35  & 3550\\
J0910.2$-$5044 & 00031282001  & 2008-10-17  & 02:18:51 & 7188\\
 & 00031282002   & 2008-10-18    & 02:25:20  & 4686\\
 & 00031282003  & 2008-11-17    & 15:13:22  & 5295\\
J1045.6$-$5937 & 00090033001  & 2009-03-24  & 08:44:01 & 14363\\
J1115.8$-$6108 & 00090051001  & 2008-04-23  & 04:50:01 & 3048\\
J1231.5$-$1410 & 00031354001  & 2009-02-24  & 18:07:41 & 4175\\
J1311.9$-$3419 & 00031358001  & 2009-02-27  & 18:24:02 & 3352\\
J1326.6$-$5302 & 00031458001  & 2009-08-05  & 14:37:45 & 4866\\
J1413.1$-$6203 & 00031410002  & 2009-05-13  & 00:35:18 & 3232\\
J1536.7$-$4947 & 00090193001  & 2009-06-20  & 02:30:01 & 1008\\
J1604.0$-$4904 & 00039227001  & 2009-05-13  & 05:03:03 & 772\\
J1634.9$-$4737 & 00312579016  & 2008-07-31  & 06:44:26 & 5285\\
J1653.4$-$0200 & 00031379001  & 2009-03-22  & 00:28:57 & 4786\\
J1746.0$-$2900 & 00035063094  & 2006-06-15  & 00:49:01 & 17528\\
J1805.3$-$2138 & 00035156001  & 2005-11-03  & 01:15:42 & 11508\\
J1813.5$-$1248 & 00031381001  & 2009-03-26  & 18:27:26 & 3232\\
 & 00031381002  & 2009-03-27    & 00:53:14  & 2314\\
 & 00031381003  & 2009-03-29    & 04:22:02  & 3974\\
& 00090197001  & 2009-06-12    & 09:30:36  & 2259\\ 
J1830.3+0617 & 00039228001  & 2009-05-20  & 04:41:20 & 580\\
J2001.0+4352 & 00039229001  & 2009-06-12  & 07:00:38 & 7334\\
J2027.5+3334 & 00090200001  & 2009-04-26  & 18:30:24 & 3490\\
J2055.5+2540 & 00031391001  & 2009-04-02  & 16:13:06 & 4749\\
J2241.7$-$5239 & 00031384001  & 2009-03-26  & 00:50:34 & 3851\\
J2302.9+4443 & 00031346001  & 2009-02-15  & 02:07:03 & 5175\\
             & 00031346002  & 2009-03-01  & 17:52:37 & 3934\\
\enddata
\end{deluxetable}
\clearpage

\begin{deluxetable}{lccccccc}
\tablewidth{0pt}
\tabletypesize{\scriptsize}
\tablecaption{0FGL J0614.3$-$3330}
\tablehead{
\colhead{Source}       & \colhead{RA}                &
\colhead{Decl.}            & \colhead{Positional error}                &
\colhead{Count rate} & HR  & $N_{\rm H}$ &\colhead{ObsID}       \\

\colhead{}                        & \colhead{\scriptsize (J2000)}         &
\colhead{\scriptsize(J2000)}     & \colhead{} &
\colhead{(0.3--10 keV)} & $\frac{H - S}{H + S}$ & (cm$^{-2}$) & \colhead{}
}
\startdata
Swift J0614.5$-$3332 & 06:14:30.0  & -33:32:22    & 6.\arcsec5 & 
$(4.5 \pm 1.1) \times 10^{-3}$ s$^{-1}$ & -0.73 & 
$3.5 \times 10^{20}$ & 00031375001\\
\enddata
\end{deluxetable}

\begin{deluxetable}{lccccccc}
\tablewidth{0pt}
\tabletypesize{\scriptsize}
\tablecaption{0FGL J0910.2$-$5044}
\tablehead{
\colhead{Source}       & \colhead{RA}                &
\colhead{Decl.}            & \colhead{Positional Error}                &
\colhead{Count rate}  & HR  & $N_{\rm H}$ & \colhead{ObsID}       \\

\colhead{}                        & \colhead{\scriptsize (J2000)}         &
\colhead{\scriptsize(J2000)}     & \colhead{} &
\colhead{(0.3--10 keV)} & $\frac{H - S}{H + S}$ & (cm$^{-2}$) & \colhead{}
}
\startdata
Swift J0910.9$-$5048 & 09:10:57.4  & -50:48:11    & 4.\arcsec6 & 
$(6.0 \pm 1.1) \times 10^{-3}$ s$^{-1}$ & 0.49 & $1.0 \times 10^{22}$& 
00031282001\\
 & 09:10:57.5  & -50:48:09    & 4.\arcsec8 &
$(1.0 \pm 0.1) \times 10^{-2}$ s$^{-1}$ & 0.72 & $1.0 \times 10^{22}$& 
00031282002\\
 & 09:10:57.7  & -50:48:05    & 4.\arcsec8 &
$(7.4 \pm 1.2) \times 10^{-3}$ s$^{-1}$ & 0.88 & $1.0 \times 10^{22}$& 
00031282003\\

\enddata
\end{deluxetable}

\begin{deluxetable}{lccccccc}
\tablewidth{0pt}
\tabletypesize{\scriptsize}
\tablecaption{0FGL J1231.5$-$1410}
\tablehead{
\colhead{Source}       & \colhead{RA}                &
\colhead{Decl.}            & \colhead{Positional error}                &
\colhead{Count rate} & HR  & $N_{\rm H}$ &  \colhead{ObsID}       \\

\colhead{}                        & \colhead{\scriptsize (J2000)}         &
\colhead{\scriptsize(J2000)}     & \colhead{} &
\colhead{(0.3--10 keV)} & $\frac{H - S}{H + S}$ & (cm$^{-2}$) & \colhead{}
}
\startdata
Swift J1231.1$-$1411 & 12:31:11.3  & -14:11:43    & 5.\arcsec6 & $(5.5 \pm 1.2) 
\times 10^{-3}$ s$^{-1}$ & -1.00 & $3.4 \times 10^{20}$ & 00031354001\\
\enddata
\end{deluxetable}

\begin{deluxetable}{lccccccc}
\tablewidth{0pt}
\tabletypesize{\scriptsize}
\tablecaption{0FGL J1311.9$-$3419}
\tablehead{
\colhead{Source}       & \colhead{RA}                &
\colhead{Decl.}            & \colhead{Positional error}                &
\colhead{Count rate} & HR  & $N_{\rm H}$ &  \colhead{ObsID}       \\

\colhead{}                        & \colhead{\scriptsize (J2000)}         &
\colhead{\scriptsize(J2000)}     & \colhead{} &
\colhead{(0.3--10 keV)} & $\frac{H - S}{H + S}$ & (cm$^{-2}$) & \colhead{}
}
\startdata
Swift J1311.5$-$3418 & 13:11:30.5  & -34:18:11    & 6.\arcsec0 & $(6.4 \pm 1.4) 
\times 10^{-3}$ s$^{-1}$ & -0.76 & $5.0 \times 10^{20}$& 00031358001\\

\enddata
\end{deluxetable}

\begin{deluxetable}{lccccccc}
\tablewidth{0pt}
\tabletypesize{\scriptsize}
\tablecaption{0FGL J1326.6$-$5302}
\tablehead{
\colhead{Source}       & \colhead{RA}                &
\colhead{Decl.}            & \colhead{Positional error}                &
\colhead{Count rate} & HR  & $N_{\rm H}$ &  \colhead{ObsID}       \\

\colhead{}                        & \colhead{\scriptsize (J2000)}         &
\colhead{\scriptsize(J2000)}     & \colhead{} &
\colhead{(0.3--10 keV)} & $\frac{H - S}{H + S}$ & (cm$^{-2}$) & \colhead{}
}
\startdata
Swift J1326.8$-$5256 & 13:26:49.4  & -52:56:26    & 3.\arcsec9 & 
$(4.2 \pm 0.3) \times 10^{-2}$ s$^{-1}$ & -0.03 & 
$1.9 \times 10^{21}$ & 00031458001\\
\enddata
\end{deluxetable}




\begin{deluxetable}{lccccccc}
\tablewidth{0pt}
\tabletypesize{\scriptsize}
\tablecaption{0FGL J1536.7$-$4947}
\tablehead{
\colhead{Source}       & \colhead{RA}                &
\colhead{Decl.}            & \colhead{Positional error}                &
\colhead{Count rate} & HR  & $N_{\rm H}$ &  \colhead{ObsID}       \\

\colhead{}                        & \colhead{\scriptsize (J2000)}         &
\colhead{\scriptsize(J2000)}     & \colhead{} &
\colhead{(0.3--10 keV)} & $\frac{H - S}{H + S}$ & (cm$^{-2}$) & \colhead{}
}
\startdata
Swift J1536.2$-$4944 & 15:36:11.7  & -49:44:57    & 7.\arcsec3 & $ (1.3 \pm
0.4) \times 10^{-2}$ s$^{-1}$ & 0.14 & $3.5 \times 10^{21}$ & 00090193001\\

\enddata
\end{deluxetable}

\begin{deluxetable}{lccccccc}
\tablewidth{0pt}
\tabletypesize{\scriptsize}
\tablecaption{0FGL J1604.0$-$4904}
\tablehead{
\colhead{Source}       & \colhead{RA}                &
\colhead{Decl.}            & \colhead{Positional error}                &
\colhead{Count rate}  & HR  & $N_{\rm H}$ & \colhead{ObsID}       \\

\colhead{}                        & \colhead{\scriptsize (J2000)}         &
\colhead{\scriptsize(J2000)}     & \colhead{} &
\colhead{(0.3--10 keV)} & $\frac{H - S}{H + S}$ & (cm$^{-2}$) & \colhead{}
}
\startdata
Swift J1603.8$-$4904 & 16:03:50.5  & -49:04:02    & 8.\arcsec6 & $ (1.0 \pm
0.4) \times 10^{-2}$ s$^{-1}$ & -0.25 & $7.9 \times 10^{21}$ & 00039227001\\

\enddata
\end{deluxetable}




\begin{deluxetable}{lccccccc}
\tablewidth{0pt}
\tabletypesize{\scriptsize}
\tablecaption{0FGL J1653.4$-$0200}
\tablehead{
\colhead{Source}       & \colhead{RA}                &
\colhead{Decl.}            & \colhead{Positional error}                &
\colhead{Count rate} & HR  & $N_{\rm H}$ &  \colhead{ObsID}       \\

\colhead{}                        & \colhead{\scriptsize (J2000)}         &
\colhead{\scriptsize(J2000)}     & \colhead{} &
\colhead{(0.3--10 keV)} & $\frac{H - S}{H + S}$ & (cm$^{-2}$) & \colhead{}
}
\startdata
Swift J1653.3$-$0158 & 16:53:15.4 & -01:58:22    & 4.\arcsec1 & $ (2.5 \pm
0.4) \times 10^{-2}$ s$^{-1}$ & -0.53 & $8.3 \times 10^{20}$ & 00031379001\\
Swift J1653.9$-$0203 & 16:53:58.7 & -02:03:15    & 5.\arcsec1 & $ (5.8 \pm     
1.1) \times 10^{-3}$ s$^{-1}$ & -0.31 & $8.3 \times 10^{20}$ &00031379001\\

\enddata
\end{deluxetable}

\begin{deluxetable}{lccccccc}
\tablewidth{0pt}
\tabletypesize{\scriptsize}
\tablecaption{0FGL J1805.3$-$2138 (Partially covered)}
\tablehead{
\colhead{Source}       & \colhead{RA}                &
\colhead{Decl.}            & \colhead{Positional error}                &
\colhead{Count rate} &  HR  & $N_{\rm H}$ & \colhead{ObsID}       \\

\colhead{}                        & \colhead{\scriptsize (J2000)}         &
\colhead{\scriptsize(J2000)}     & \colhead{} &
\colhead{(0.3--10 keV)} & $\frac{H - S}{H + S}$ & (cm$^{-2}$) & \colhead{}
}
\startdata
Swift J1804.5$-$2140 & 18:04:32.3 & -21:40:10    & 4.\arcsec6 & $ (1.8 \pm
0.5) \times 10^{-3}$ s$^{-1}$ & 1.00 & $1.1 \times 10^{22}$ &00035156001\\

\enddata
\end{deluxetable}

\begin{deluxetable}{lccccccc}
\tablewidth{0pt}
\tabletypesize{\scriptsize}
\tablecaption{0FGL J1813.5$-$1248}
\tablehead{
\colhead{Source}       & \colhead{RA}                &
\colhead{Decl.}            & \colhead{Positional error}                &
\colhead{Count rate}   & HR  & $N_{\rm H}$ & \colhead{ObsID}       \\

\colhead{}                        & \colhead{\scriptsize (J2000)}         &
\colhead{\scriptsize(J2000)}     & \colhead{} &
\colhead{(0.3--10 keV)} & $\frac{H - S}{H + S}$ & (cm$^{-2}$) & \colhead{}
}
\startdata
Swift J1813.4$-$1246 & 18:13:23.5  & -12:46:00    & 5.\arcsec0 & 
$(8.1 \pm 1.8) \times 10^{-3}$ s$^{-1}$ & 0.96
& $6.6 \times 10^{21}$ & 00031381001\\
 & 18:13:23.5  & -12:46:03    & 7.\arcsec1 &
$(5.2 \pm 1.6) \times 10^{-3}$ s$^{-1}$ & 0.50 
&$6.6 \times 10^{21}$ & 00031381002\\
 & 18:13:23.3  & -12:46:01    & 4.\arcsec8 &
$(7.2 \pm 1.5) \times 10^{-3}$ s$^{-1}$ & 1.00 &$6.6 \times 10^{21}$ & 00031381003\\
 & 18:13:23.6  & -12:46:04    & 6.\arcsec0 &
$(7.5 \pm 1.5) \times 10^{-3}$ s$^{-1}$ & 0.50 
&$6.6 \times 10^{21}$ & 00090197001\\

\enddata
\end{deluxetable}

\begin{deluxetable}{lccccccc}
\tablewidth{0pt}
\tabletypesize{\scriptsize}
\tablecaption{0FGL J1830.3+0617}
\tablehead{
\colhead{Source}       & \colhead{RA}                &
\colhead{Decl.}            & \colhead{Positional error}                &
\colhead{Count rate} &  HR  & $N_{\rm H}$ &\colhead{ObsID}       \\

\colhead{}                        & \colhead{\scriptsize (J2000)}         &
\colhead{\scriptsize(J2000)}     & \colhead{} &
\colhead{(0.3--10 keV)} & $\frac{H - S}{H + S}$ & (cm$^{-2}$) & \colhead{}
}
\startdata
Swift J1830.1+0619 & 18:30:05.8 & 06:19:12    & 6.\arcsec0 & $ (5.3 \pm
0.8) \times 10^{-2}$ s$^{-1}$ & 0.2 & $2.4 \times 10^{21}$ & 00039228001\\

\enddata
\end{deluxetable}

\begin{deluxetable}{lccccccc}
\tablewidth{0pt}
\tabletypesize{\scriptsize}
\tablecaption{0FGL J2001.0+4352}
\tablehead{
\colhead{Source}       & \colhead{RA}                &
\colhead{Decl.}            & \colhead{Positional error}                &
\colhead{Count rate} &  HR  & $N_{\rm H}$ &\colhead{ObsID}       \\

\colhead{}                        & \colhead{\scriptsize (J2000)}         &
\colhead{\scriptsize(J2000)}     & \colhead{} &
\colhead{(0.3--10 keV)} & $\frac{H - S}{H + S}$ & (cm$^{-2}$) & \colhead{}
}
\startdata
Swift J2001.1+4348 & 20:01:03.6 & 43:48:29    & 5.\arcsec8 & $ (1.8 \pm       
0.3) \times 10^{-3}$ s$^{-1}$ & -0.08 &$3.7 \times 10^{21}$ &00039229001\\
Swift J2001.2+4352 & 20:01:12.7 & 43:52:49    & 3.\arcsec7 & $ (4.8 \pm
0.3) \times 10^{-2}$ s$^{-1}$ & -0.54 &$3.7 \times 10^{21}$ & 00039229001\\
\enddata
\end{deluxetable}




\begin{deluxetable}{lccccccc}
\tablewidth{0pt}
\tabletypesize{\scriptsize}
\tablecaption{0FGL J2055.5+2540}
\tablehead{
\colhead{Source}       & \colhead{RA}                &
\colhead{Decl.}            & \colhead{Positional error}                &
\colhead{Count rate} &  HR  & $N_{\rm H}$ &\colhead{ObsID}       \\

\colhead{}                        & \colhead{\scriptsize (J2000)}         &
\colhead{\scriptsize(J2000)}     & \colhead{} &
\colhead{(0.3--10 keV)} & $\frac{H - S}{H + S}$ & (cm$^{-2}$) & \colhead{}
}
\startdata
Swift J2055.8+2546 & 20:55:48.5 & 25:46:35    & 5.\arcsec4 & $ (4.0 \pm
1.0) \times 10^{-3}$ s$^{-1}$ & -0.05 & $1.1 \times 10^{21}$ & 00031391001\\
Swift J2055.8+2540 & 20:55:50.3 & 25:40:49    & 6.\arcsec0 & $ (3.2 \pm     
1.1) \times 10^{-3}$ s$^{-1}$ & 0.07 & $1.1 \times 10^{21}$ & 00031391001\\

\enddata
\end{deluxetable}

\begin{deluxetable}{lccccccc}
\tablewidth{0pt}
\tabletypesize{\scriptsize}
\tablecaption{0FGL J2241.7$-$5239}
\tablehead{
\colhead{Source}       & \colhead{RA}                &
\colhead{Decl.}            & \colhead{Positional error}                &
\colhead{Count rate} & HR  & $N_{\rm H}$ & \colhead{ObsID}       \\

\colhead{}                        & \colhead{\scriptsize (J2000)}         &
\colhead{\scriptsize(J2000)}     & \colhead{} &
\colhead{(0.3--10 keV)} & $\frac{H - S}{H + S}$ & (cm$^{-2}$) & \colhead{}
}
\startdata
Swift J2241.5$-$5246 & 22:41:33.3 & -52:46:34    & 5.\arcsec4 & $ (5.4 \pm
1.3) \times 10^{-3}$ s$^{-1}$ & -0.33 & $1.2 \times 10^{20}$ & 00031384001\\

\enddata
\end{deluxetable}

\begin{deluxetable}{lccccccc}
\tablewidth{0pt}
\tabletypesize{\scriptsize}
\tablecaption{0FGL J2302.9+4443}
\tablehead{
\colhead{Source}       & \colhead{RA}                &
\colhead{Decl.}            & \colhead{Positional error}                &
\colhead{Count rate} &  HR  & $N_{\rm H}$ & \colhead{ObsID}       \\

\colhead{}                        & \colhead{\scriptsize (J2000)}         &
\colhead{\scriptsize(J2000)}     & \colhead{} &
\colhead{(0.3--10 keV)} & $\frac{H - S}{H + S}$ & (cm$^{-2}$) & \colhead{}
}
\startdata
Swift J2302.1+4445 & 23:02:08.7 & 44:45:33    & 6.\arcsec7 & $ (2.3 \pm
0.8) \times 10^{-3}$ s$^{-1}$ & -0.83 & $1.3 \times 10^{21}$ & 00031346001\\
         &  23:02:08.7  & 44:45:33    & -- & $ < 8.0 
          \times 10^{-4}$ s$^{-1}$ & --  & $1.3 \times 10^{21}$ & 00031346002\\

\enddata
\end{deluxetable}

\begin{figure}[t]
\hfil
\includegraphics[width=1.\linewidth]{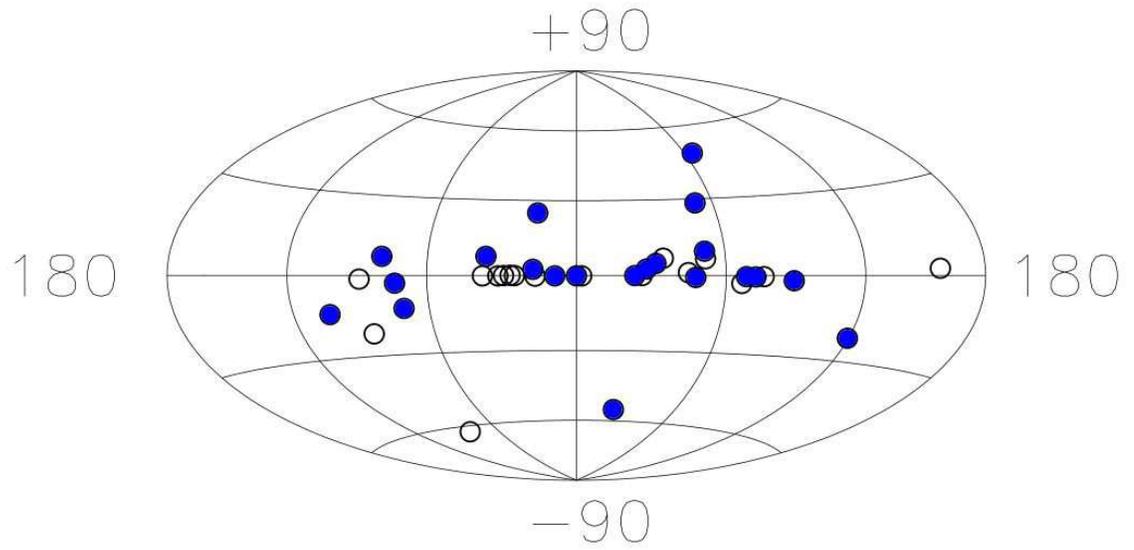}
\hfil
\caption{
Locations of unidentified 0FGL sources listed in \citet{abdo1}. 
Shaded symbols indicate sources observed by \swift. 
}
\label{radio}
\end{figure}

\end{document}